\documentclass[lettersize,journal]{IEEEtran}
\usepackage{amsmath,amsfonts}
\usepackage{algorithmic}
\usepackage{algorithm}
\usepackage{array}
\usepackage{authblk}
\usepackage[caption=false,font=normalsize,labelfont=sf,textfont=sf]{subfig}
\usepackage{textcomp}
\usepackage{stfloats}
\usepackage{url}
\usepackage{verbatim}
\usepackage{graphicx}
\usepackage{cite}
\usepackage[dvipsnames]{xcolor}

\begin{document}

\title{TES Bolometer Design and Testing for the \\ Tomographic Ionized-carbon Mapping Experiment Millimeter Array}

\author[a]{Victoria L. Butler}
\author[b]{James J. Bock}
\author[c,a]{Dongwoo T. Chung}
\author[a,b]{Abigail T. Crites}
\author[b]{Clifford Frez}
\author[b]{King Lau}
\author[d]{Ian Lowe}
\author[d]{Dan P. Marrone}
\author[d]{Evan C. Mayer}
\author[a]{Benjamin J. Vaughan}
\author[e]{Michael Zemcov}
\affil[a]{Cornell University, Ithaca, NY, U.S.A.}
\affil[b]{California Institute of Technology, Pasadena, CA, U.S.A.}
\affil[c]{University of Toronto, Toronto, Ontario, Canada}
\affil[d]{University of Arizona, Tucson, AZ, U.S.A.}
\affil[e]{Rochester Institute of Technology, Rochester, NY, U.S.A.}

\maketitle

\begin{abstract}
Transition Edge Sensor (TES) bolometers are a well-established technology with a strong track record in experimental cosmology, making them ideal for current and future radio astronomy instruments. The Tomographic Ionized-carbon Mapping Experiment (TIME), in collaboration with JPL, has developed advanced silicon-nitride leg-isolated superconducting titanium detectors for 200–300 GHz observations of the Epoch of Reionization. Compared to their MHz counterparts, bolometers operating in this frequency range are less common because of their large absorber size and fragility. TIME aims to fabricate a total of 1920 high-frequency (HF) and low-frequency (LF) detectors to fully populate the focal plane.

TIME has successfully developed HF ($\sim 230-325$\ GHz) and LF ($\sim 183-230$ GHz) wafers that are physically robust and perform well at cryogenic temperatures ($\sim $300 mK). Recent laboratory tests have shown high optical efficiencies for the LF wafers (30-40\%), but low device yield for the HFs. To address this, new HF modules have been designed with improved cabling and a reduced backshort distance, and are expected to perform similarly to LFs in a similar lab setting. We report on the development of these detectors as well as recent laboratory and on-sky tests conducted at the Arizona Radio Observatory's (ARO) 12-meter prototype antenna at Kitt Peak National Observatory.
\end{abstract}

\begin{IEEEkeywords}
transition-edge sensor bolometers, line intensity mapping, spectroscopy, mm-wavelength
\end{IEEEkeywords}

\section{TIME LIM Science}
    
    The Tomographic Ionized-carbon Mapping Experiment (TIME \cite{crites14}) is designed to observe [C\textsc{\,ii}] and CO rotation transition emission lines across multiple high redshifts using the technique of line intensity mapping (LIM) \cite{LIMreview17,LIMreview19,LIMreview22}. With sensitivity to [C\textsc{\,ii}] emission at roughly $z\approx5$--9, TIME will measure the dust-obscured star formation history in a period during reionization \cite{Sun21,looze2014}. This will bridge the knowledge gap between the earliest star formation processes and the fully ionized galactic universe that existed at Cosmic Noon. LIM has the potential to independently confirm various cosmological models, as well as provide cross correlation with other tracers and instruments (see e.g. FYST \cite{fyst_lim}, CONCERTO \cite{concerto_lim}, mmIME \cite{mmime_lim},COMAP \cite{comap_lim}, EXCLAIM \cite{exclaim_lim}, HETDEX \cite{hetdex_lim}, SPHEREx \cite{spherex_lim}). TIME will also be sensitive to low-to-mid CO rotational transitions between $z=0.5$--2, revealing the molecular gas history during the epoch of peak star formation \cite{keating2016,keating2020}.
    

    \begin{figure}
        \centering
        \includegraphics[width=\linewidth]{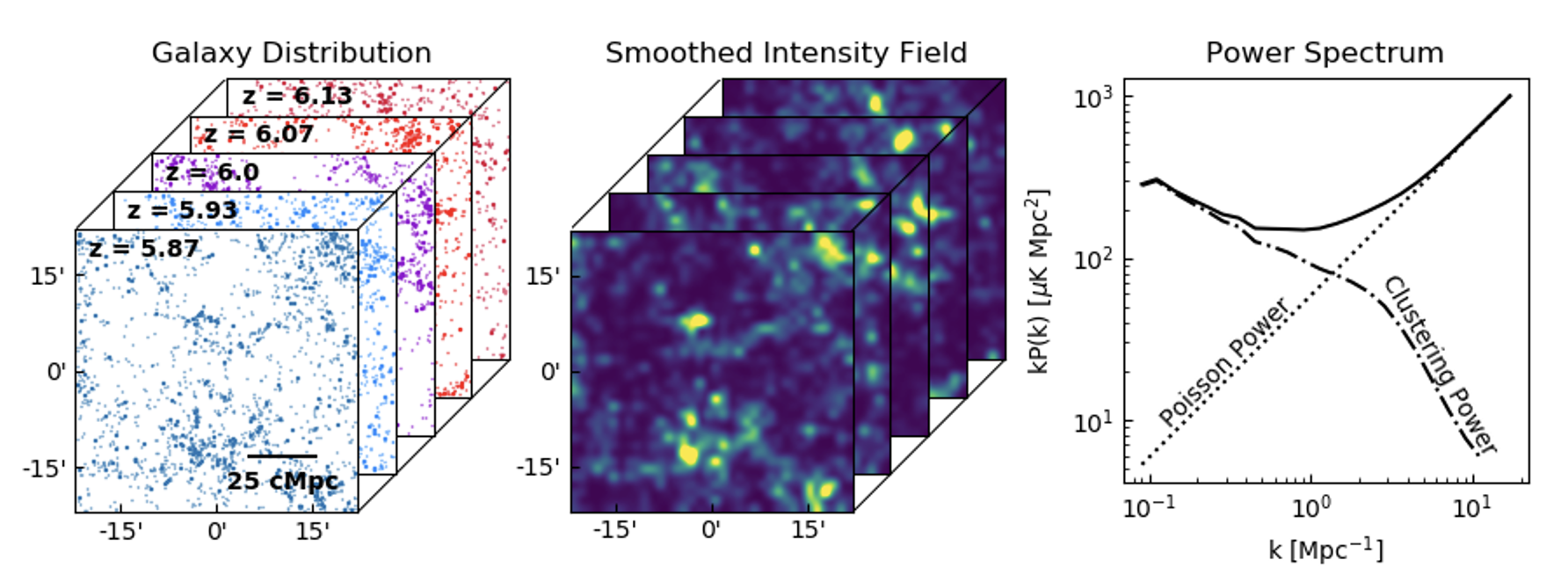}
        \caption{TIME will perform LIM of the Epoch of Reionization and the Peak of Cosmic Star Formation ($z\sim2$). LIM provides three dimensional information about the evolution of [C\textsc{\,ii}] and CO emission lines over time, revealing the growth of structure. The left image shows a simulation of individual galaxies at multiple redshifts, while the middle image shows the estimated TIME intensity map from these sources using a 30$''$ beam, multiple frequency channels, and no measurement noise. The right figure shows the power spectrum from the left image and the expected contributions from clustering and Poisson fluctuations. 
        }
        \label{fig:enter-label}
    \end{figure}
    
    The enabling technology for CO/[C\textsc{\,ii}] LIM with TIME are transition-edge sensor (TES) Bolometers, which have a rich heritage in past Cosmology experiments \cite{tes}. Combined with modern broadband spectrometers sensitive to both polarizations \cite{Bradford,Li2018}, and an optimized field of view of 1.3$^\circ$~$\times$~0.45$^\prime$, TIME will maximize its $S/N$ of baryon clustering.
    

\section{Instrument Overview}

    \begin{figure*}
        \centering
        \includegraphics[width=\linewidth]{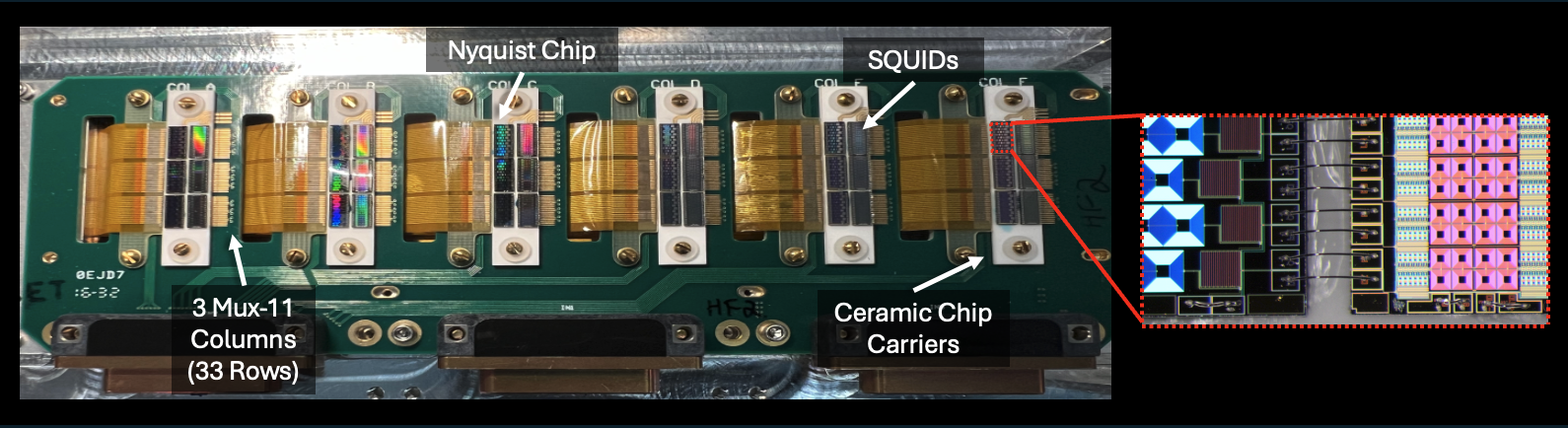}
        \caption{Amplifier SQUID module connected to the detector side through Kapton\texttrademark\ flex cables. The TIME architecture has 3 subarrays of Nyquist and SQUID chips (6 total), group in packets of 11 multiplexed rows for a total of 33. On the right is a close-up of a single Nyquist and SQUID chip. The Nyquist chips were designed and fabricated at NIST.}
        \label{fig:squids}
    \end{figure*}

The TIME instrument is a multistage cryostat designed to operate at a base temperature of 250 mK, necessary for our superconducting titanium detectors. This base stage is achieved through the use of a Cryomech PT415 cryocooler that reaches 50K and 4K, a $^{4}$He Joule-Thomson fridge to maintain 1K, and a final series of $^{3}$He sorption fridges to reach milli-Kelvin temperatures. 

\begin{figure}
        \centering
        \includegraphics[width=\linewidth]{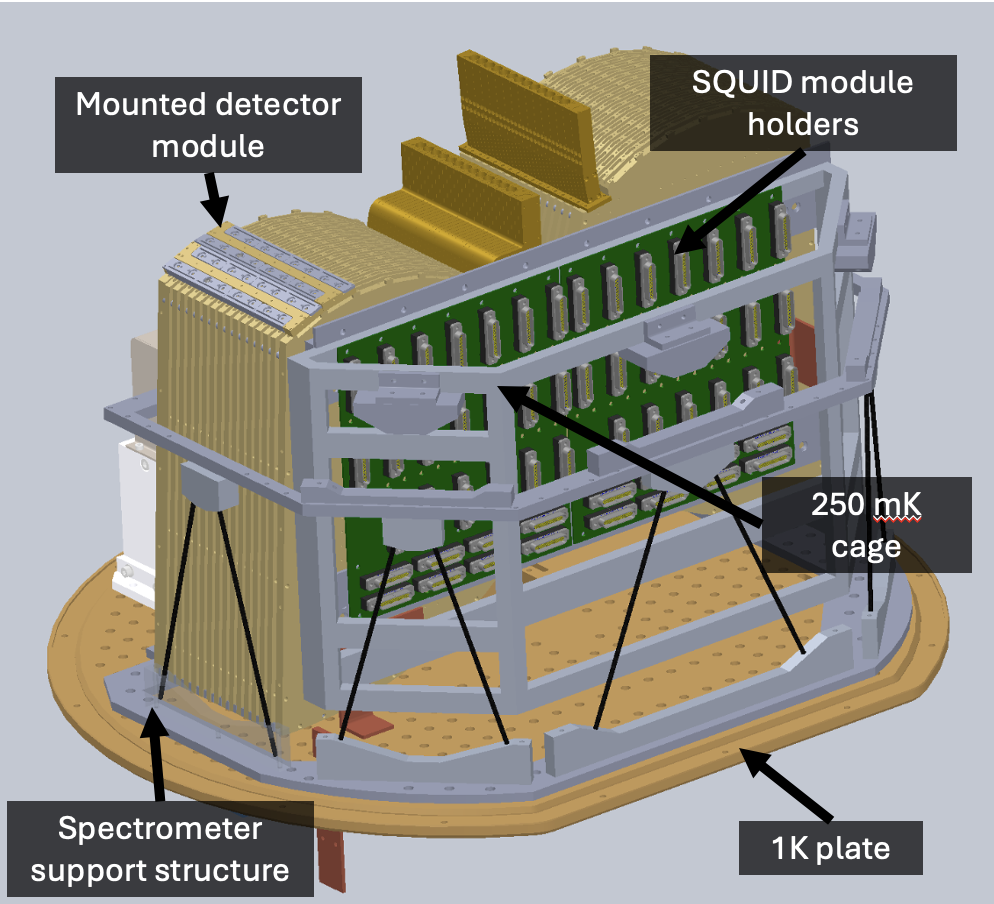}
        \caption{The TIME spectrometers mounted to the 1K plate, and held in place using a rigid stage assembly. During installation, the feedhorns point towards the ground.}
        \label{fig:spec}
    \end{figure}

\begin{figure}
        \centering
        \includegraphics[width=\linewidth]{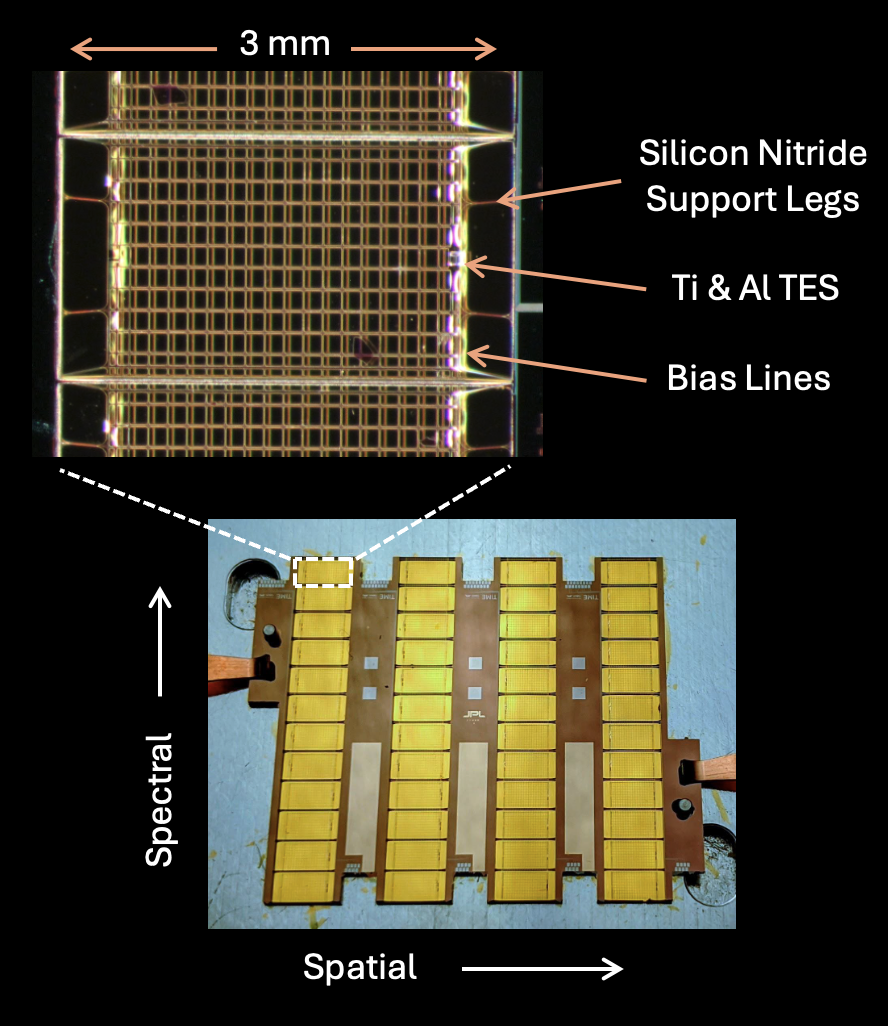}
        \caption{The bottom figure shows an individual HF subarray, 4 of which are arranged in parallel to form a single HF module. The number of detectors varies across the high and low frequency subarrays. This HF wafer was designed by Jonathan Hunacek and created at the JPL microdevices laboratory by Clifford Frez and Anthony Turner. Above is a close-up of an individual TES Bolometer comprising a gold absorber, Ti and AL TES, and silicon nitride support legs which suspend the absorber.}
        \label{fig:modules}
    \end{figure}

The milli-Kelvin stage houses 1920 silicon-nitride leg-isolated detectors arrayed across two gold-plated R$\sim$170 spectrometer banks. Figure~\ref{fig:spec} shows the spectral dimension along which the detector arrays are mounted, along with the 16 feedhorns that couple incident light to the detectors. There are a total of 6 detector modules mounted to each spectrometer, each with 4 subarrays, one of which is shown in Figure~\ref{fig:modules}. These modules have two different designs which are optimized for different portions of our $183-325$ GHz window. The designed properties of these two types are listed in Table~\ref{tab:det} where LF refers to ``Low Frequency" and HF to ``High Frequency". An individual TIME LF absorbing web is shown in the upper portion of Figure~\ref{fig:modules}, which has a slightly larger absorbing area compared to the HF wafers. The overall size of each module is the same, leading to a different number of total HF and LF pixels. 

Each module also comes with its own signal pre-amplifiers, time-domain multiplexed (TDM) Superconducting Quantum Inference Devices (SQUID)\cite{squid1,squid2}, shown in Figure \ref{fig:squids}. The TDM wiring configuration allows us to readout the detectors across 33 rows and 32 columns, which is digitized by the multi-channel electronic (MCE) crates developed by UBC \cite{MCE}. For our roughly 2000 detectors, this requires 8 readout cards spread across two MCE crates.

    \begin{table*}[t]
    \caption{A partial summary of TIME instrument specifications} 
    \label{tab:det}
    \label{tab:specspecs}
    \begin{center}       
    \begin{tabular}{|l|l|l|} 
    \hline
    \rule[-1ex]{0pt}{3.5ex}  Parameter & LF & HF  \\
    \hline
    \rule[-1ex]{0pt}{3.5ex}  Spectral range (GHz) & 183--230 & 230--325   \\
    \hline
    \rule[-1ex]{0pt}{3.5ex}  End-to-end optical efficiency & 0.2--0.25 & 0.2--0.25  \\
    \hline
    \rule[-1ex]{0pt}{3.5ex}  Resolving power $\nu/\delta\nu$ & 90--120 & 90--120 \\
    \hline
    \rule[-1ex]{0pt}{3.5ex}  Bolometer count / subarray & 32 (1 pol. $\times$ 4 feeds $\times$ 8 chan.) & 48 (1 pol. $\times$ 4 feeds $\times$ 12 chan.) \\
    \hline 
    \rule[-1ex]{0pt}{3.5ex}  Bolometer count (total) & 768 (2 pol. $\times$ 16 feeds $\times$ 24 chan.) & 1152 (2 pol. $\times$ 16 feeds $\times$ 36 chan.) \\
    \hline 
    \end{tabular}
    \end{center}
    \end{table*}

\section{Lab Detector Testing}

    There are several properties of TES bolometers used to characterize their ability to detect photons. TIME detectors are optimized for radio wavelength, extremely low energy photon absorption, which requires operation at cryogenic temperatures. Since bolometers are effectively thermometers, their performance depends on their weak thermal conductance $G$ to the surrounding bath temperature $T_{bath}$. We characterize the Ti superconducting transition temperature of the detectors as $T_{C}$, with the temperature gradient of these two temperatures setting the saturation power limit of the device, $P_{sat}$. In the sections below, we test the $P_{sat}$ levels of our newest devices. First, we discuss strategies that maximize the performance of the focal plane for these tests. 

    \subsection{Loadcurves}

    TES Bolometers operate on a property called electrothermal feedback, meaning that changes in the incident power cause temperature changes, and subsequently change the resistance of the circuit. When a constant voltage bias is applied to the circuit, this feedback can be characterized by $\frac{dP_{optical}}{dI}$. The region where this feedback, or loop gain, is the most efficient is at the superconducting transition.

\begin{figure}
        \centering
        \includegraphics[width=\linewidth]{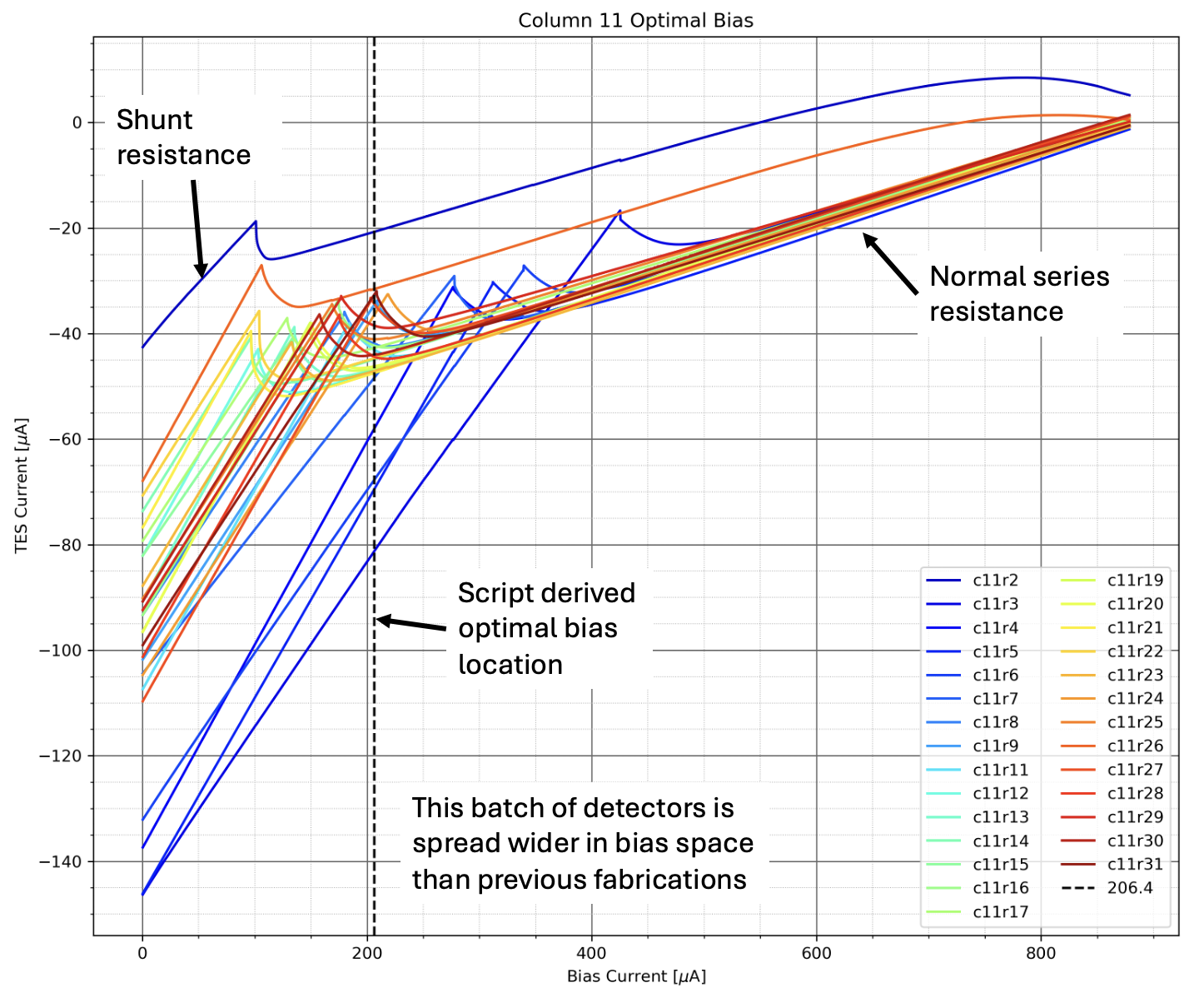}
        \caption{The bias is swept from high to low values for detectors within the same multiplexing column. Fabrication differences sometimes create detectors with diverse optimal transition bias locations. This can be a challenge for preserving  large numbers of functional detectors at a single bias current.}
        \label{fig:loadcurve}
    \end{figure}

    The bias voltage (or current) for this transition shifts with bath temperature, as well as small resistance changes from mechanical defects or readout electronics. To determine the operating behavior for the detectors, we can take what are called \textit{loadcurves}, where the bias current is stepped down from high to low at a fixed bath temperature. The detector response, shown in Figure~\ref{fig:loadcurve}, reveals the optimal transition location for a multiplexed row. Bias currents too high or low lock the detectors in a ``normal" or ``superconducting" state respectively, where the current resistance is not sensitive to incident photons. When the TES is locked in a transition state, the TES constant voltage is maintained by transferring the current through the shunt resistor ($3-4~m\Omega$). The bath temperature and bias current however must be increased significantly in order to move the TES to their transition state when the detectors reach a ``cold branch". 
    
    
    To achieve the maximum number of working detectors per multiplexed row, we have mimicked the tuning setup used for BICEP3 dark detectors. This involves ``unlatching" at an extremely high bias, 32,000 DAC\footnote{DAC are arbitrary current units generated by the MCE. These are later converted into amperes.}. We then heat the stage above the titanium superconducting normal transition (500 mK) and start the loadcurves at a high bias to avoid the cold branch, around 3,000 DAC. This is well above the unusually high optimal biases of some of our modules, which can reach nearly 2,200 DAC.

    \subsection{Saturation Power}

\begin{figure}
        \centering
        \includegraphics[width=\linewidth]{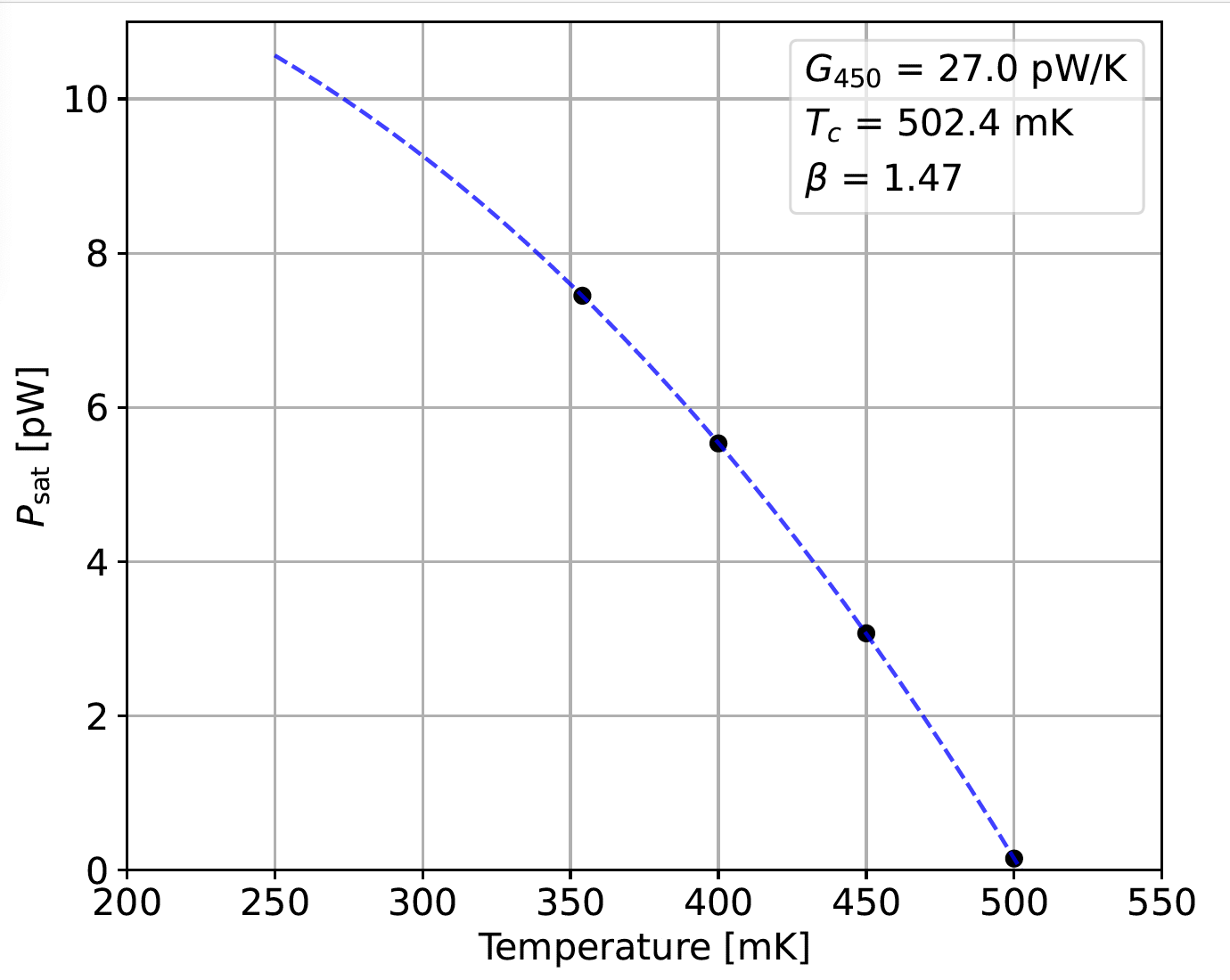}
        \includegraphics[width=\linewidth]{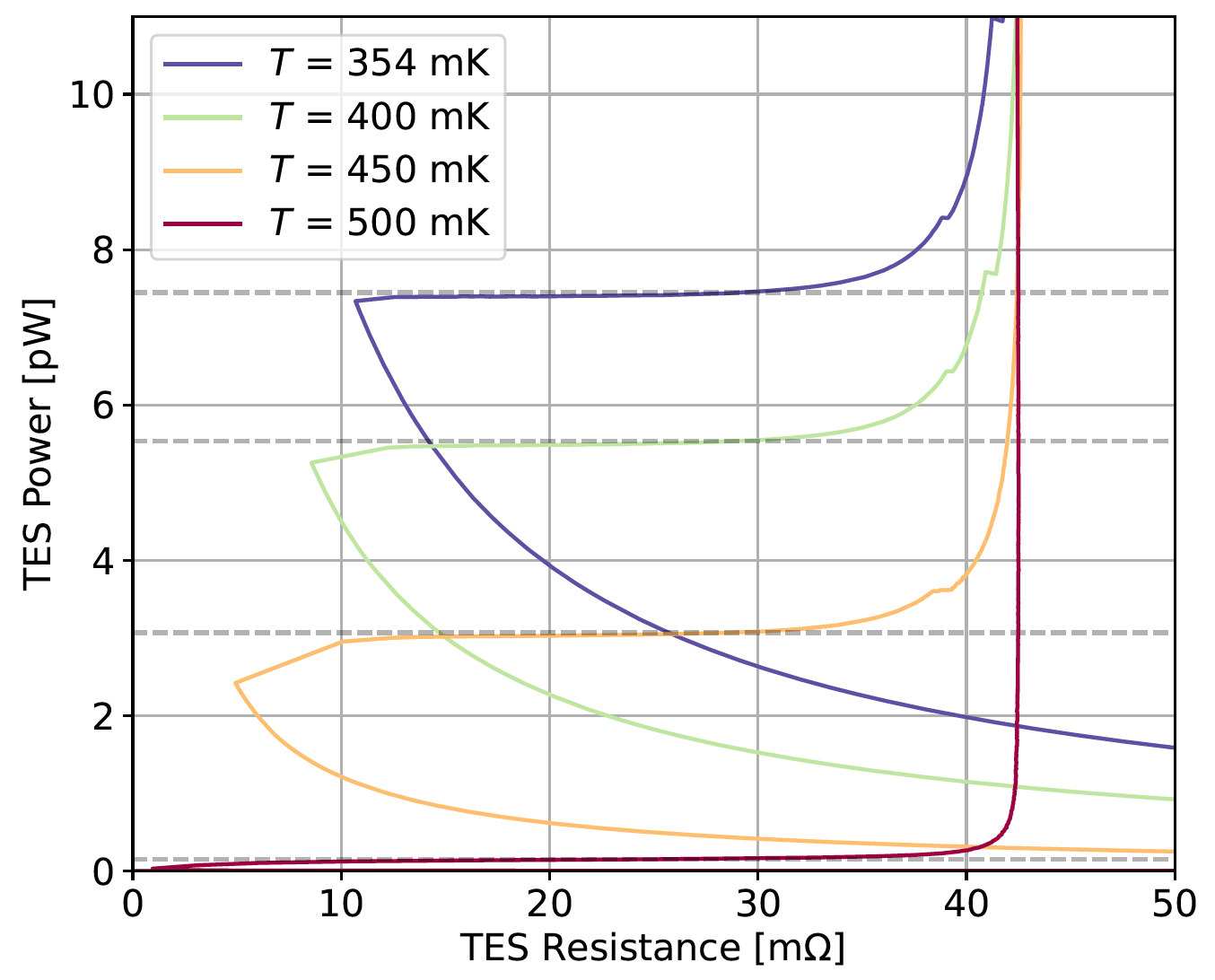}
        \caption{\textit{Top}: Values of $G$ estimated from saturation power measurements at various bath temperatures. The derived parameters shown in the top right are based on a theoretical best fit to the blue dashed line. \textit{Bottom}: $P_{sat}$ derived from various bath temperatures for a single detector. Horizontal dashed lines are estimated $P_{sat}$s for the detector's known superconducting transition.}
        \label{fig:g_values}
    \end{figure}

    Based on the geometry of our detectors, the ideal value of the thermal conductance $G$ between the bolometers and $T_{bath}$ would fall around 20~pW/K for LF arrays and 30~pW/K for HF arrays at $T_{bath}=$~450~mK. These design targets are optimized to achieve a photon noise level lower than the on-sky loading at our observation site as well as a desirable detector time constant for our science mission. As demonstrated in Figure~\ref{fig:g_values}, we determined $G$ values by measuring the saturation power $P_\text{sat}$ of each detector at multiple $T_\text{bath}$. The relationship fit to the data in the top panel of Figure \ref{fig:g_values} is represented by
    \begin{equation}\label{eq:g_calc}
        P_{\text{sat}} = \frac{G_c}{\beta + 1}\frac{T_{c}^{\beta+1}-T_{\text{bath}}^{\beta+1}}{T_{c}^{\beta}},
    \end{equation}
    where $T_c$ is the critical temperature and $G_{c} \equiv G(T=T_c)$. We then derived $G_{450}$ by the power-law dependence $G(T) \propto T^{\beta}$.
    
    In general, our results show that LF bolometers have a fairly uniform $G$ distribution with a median that hits our design target. Although the HF bolometers also show a similar property, we were able to identify several subarrays with thermal shorts as a result of fabrication errors. We are using this information to improve the fabrication process for the newest batches of detectors. 

    \subsection{Optical Efficiency}

\begin{figure}[t]
        \centering
        \includegraphics[width=\linewidth]{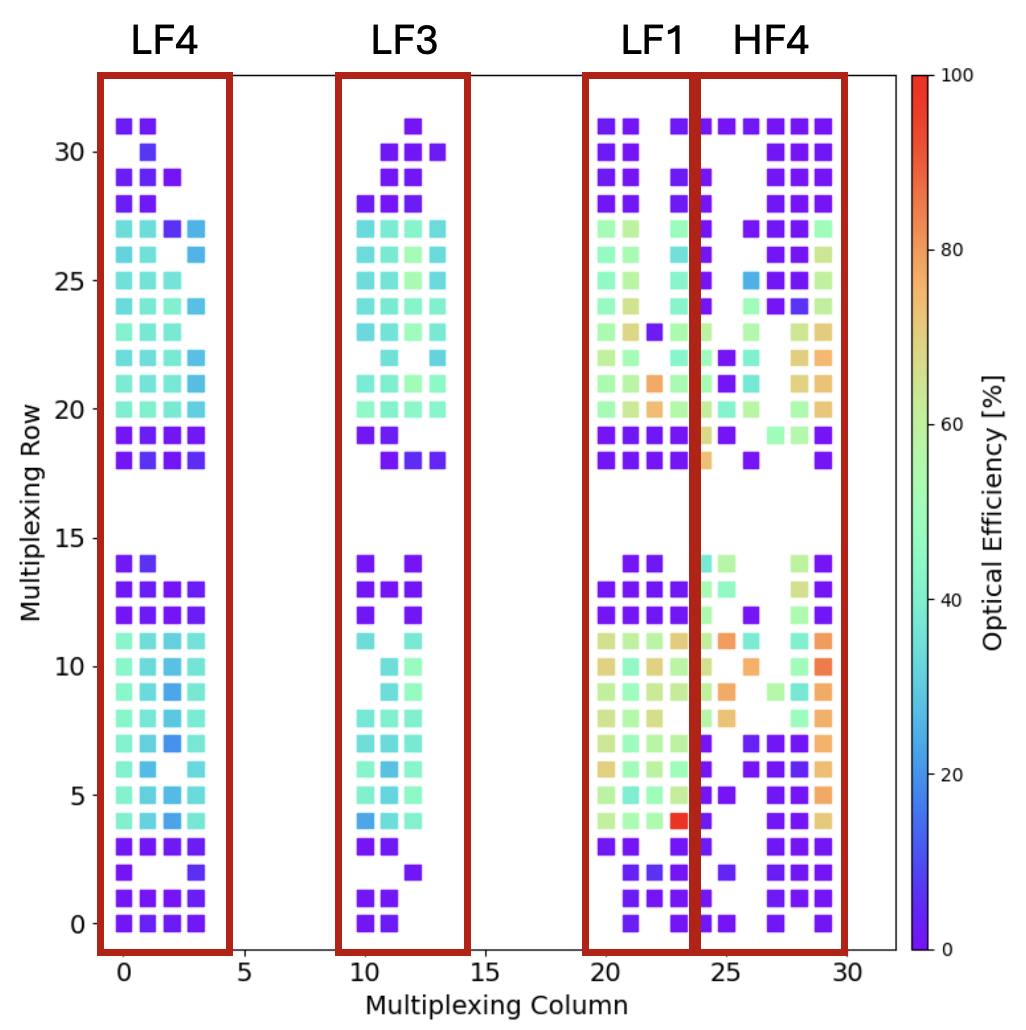}
        \caption{Optical efficiencies for 4 individual detector modules, three of which were low frequency, and one of which was high frequency. We plot individual detector responses in each cube, and identifying their position based on multiplexing readout coordinates. The results were within the expected range, however, we noted that changes in bath temperature and the quality of the thermal coupling to the 250 mK stage significantly impacted these results.}
        \label{fig:opt-eff}
    \end{figure}

    We calculated the optical efficiencies $\eta$ of the detectors using the change in power between two loading conditions defined by 
    \begin{equation}
        \eta = \frac{P_{\text{sat},77\text{K}} - P_{\text{sat},300\text{K}}}{(300\text{K} - 77\text{K}) \, k_B \Delta \nu} \, , 
    \end{equation}
    where $\Delta \nu$ is the effective bandwidth of the detector channel, which we have previously characterized as $\sim 1.5$~GHz for nearly all detectors. Our two loading cases were Eccosorb\texttrademark\ layers immersed in liquid Nitrogen with an approximate temperature of 77~K, and room temperature at roughly 300~K. We ran this test on a recent batch of detector arrays which resulted in $\eta$ values between 30--40$\%$ up to $\nu \sim 260$~GHz, shown in Figure~\ref{fig:opt-eff}. With a new HF module design that yields a smaller backshort distance (explained further in Section~\ref{sec:hf}), we expect $\eta$ to achieve the same values in the high frequency regime before gradually decreasing to zero at the end of the atmospheric window $\sim 320$~GHz, due to the low-pass filter employed above the focal plane.

\section{Module Upgrades}

    The TIME instrument has undergone years of lab testing, along with several engineering deployments at a ``high and dry" 12 meter radio telescope in Arizona. During this time, our knowledge of the functional behavior of our unique bolometer design has advanced to a stage where methodical design iteration is achievable. Recent notable changes have been the redesign and fabrication of our high frequency detectors, along with their associated readout cables. The majority of this work has been completed by collaborators at Caltech, with support from NASA's Jet Propulsion Laboratory. We have been able to perform rapid prototyping and testing of new devices through the use of a 100mK cryostat, which provides detailed insight into superconducting behavior not observable otherwise. This cryostat is small, with a much faster turn-around time from warm to cold temperatures compared to the primary TIME cryostat, located at Cornell University.

    \subsection{Kapton\texttrademark\ Cable Fabrication}
    Results from cryogenic testing conditions showed several detector modules with high series impedance, traced to the Kapton\texttrademark\ cable resistance that connects the SQUID chips and detector module electrical chain. Ideally, these cables should remain well below the superconducting resistance of the TES, which is on average between $20-50~\textrm{m}\Omega$. The original design was a 12" long by 1" wide polyamide flex cable with 0.25 oz thick tinned copper traces, separate by 5 mil spacing. The ends were coated in gold in order to facilitate adhesion with aluminum wire bonds, connecting the cable to either the detector or SQUID side of the module. This resulted in an average cable performance of 10 m$\Omega$ per wire pair (readout and bias lines) at $T_{bath}= 250$mK. This was further reduced by adding additional tin to the cable edges and masking more of the bond pad area. The result was resistances closer to 6.6m$\Omega$. Unfortunately, this design was still too high of a resistance for optimal detector performance. 

    \begin{figure}
        \centering
        \includegraphics[width=\linewidth]{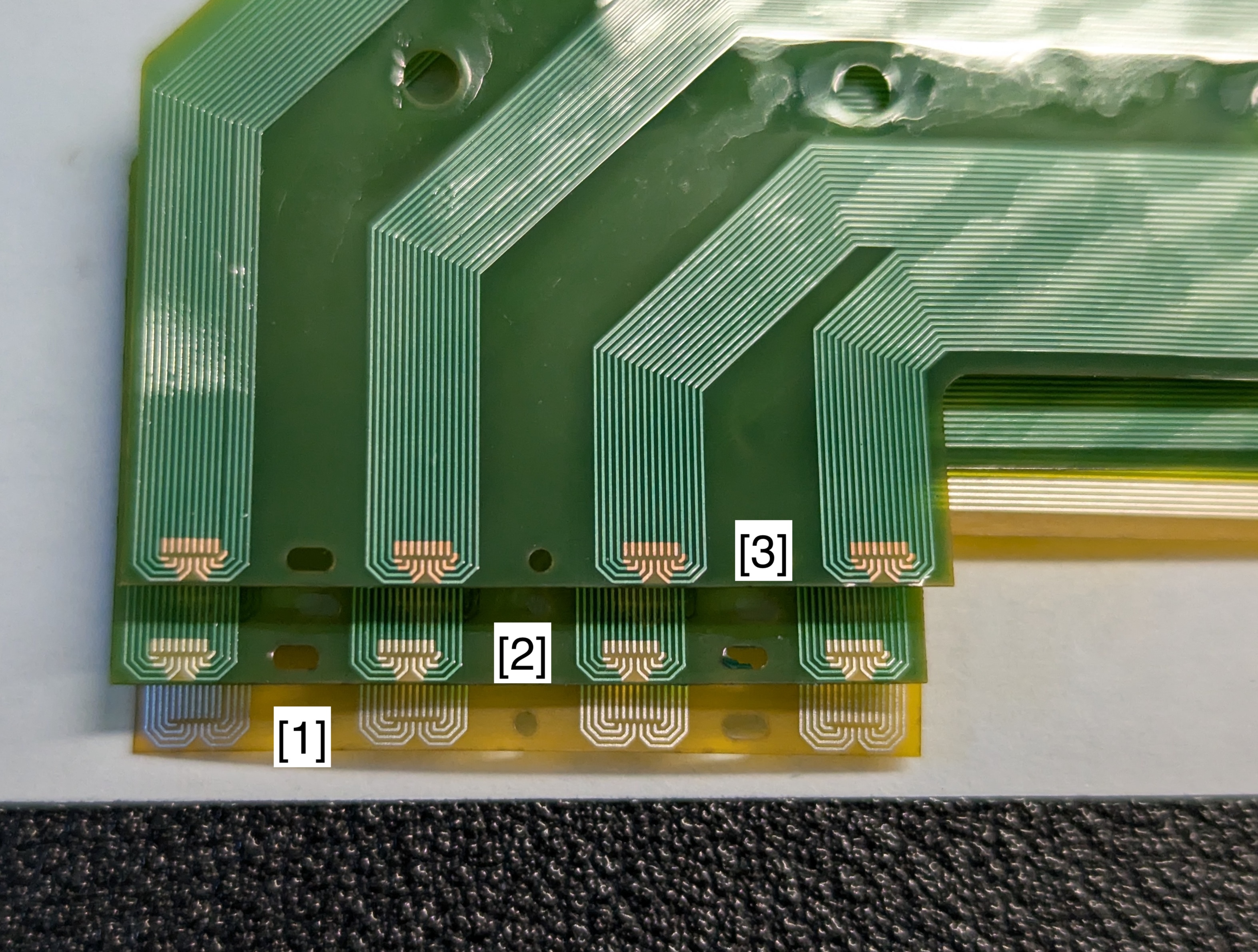}
        \caption{Newly fabricated Kapton\texttrademark\ cables showing different phases of the wet etch process. The opposite end of this cable has a plastic stiffener used to prevent tearing of the cable at the interface with the silicon SQUID board.}
        \label{fig:new_kapton}
    \end{figure}

\begin{figure*}[t]
        \centering
        \includegraphics[width=\linewidth]{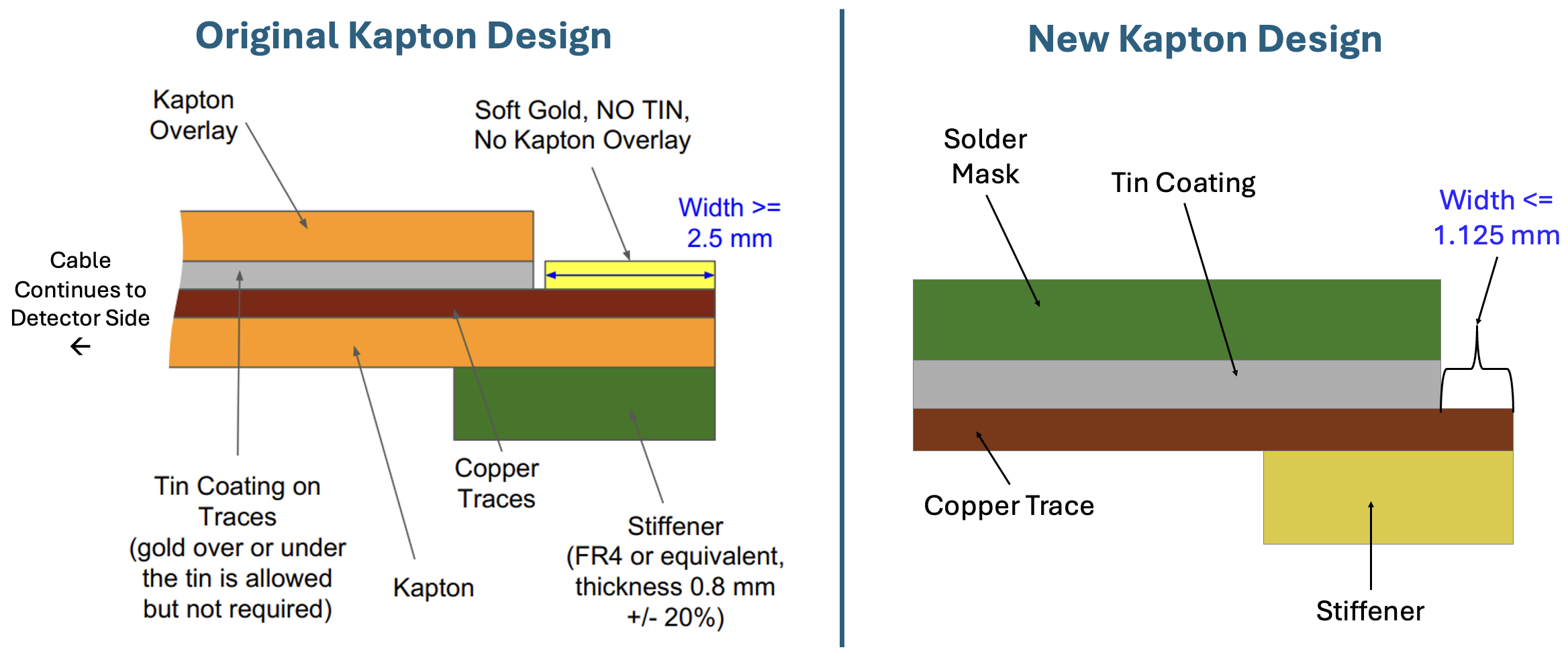}
        \caption{The original Kapton\texttrademark\ design, shown on the left, had a large bonding surface and gold plating. This is the key difference when compared to the new design. Figure recreated with permission from the author\cite{Hunacek2020}.}
        \label{fig:old_kapton}
    \end{figure*}

    The Kapton\texttrademark\ cables are undergoing a re-fabrication, which aims to improve this series impedance through reduction of the bonding area size, and replacing some of the copper with tin. The total exposed bonding pad is a fraction of the original size (down to 1.13mm from 2.5mm as shown in Figure \ref{fig:old_kapton}), resulting in a lower impedance at low temperatures. The cables at various steps of the new etching process are shown to the left, from bottom to top in Figure \ref{fig:new_kapton}. Stage [1] has only the raw tin coated wires with no gold coating, and no Kapton\texttrademark\ overlay. Stage [2] adds a green solder mask, while Stage [3] uses a \texttt{NPS 220} and \texttt{NPS 240} solution to remove the tin coating on the last $1.125$mm of exposed wire for wirebonding.

    Cryogenic testing of the remade cables showed tin superconducting transitions at $\sim$3K only for samples using the 0.25 oz without a soldermask. This suggests that either the tin deposition is insufficient/improper on thicker copper traces, or that the masking process promotes tin oxidation. We are now testing a fabrication variation where the 0.25 oz copper is combined with the soldermask with pending results. In parallel, we are also exploring alternatives such as full-cable fabrication via niobium lithography.  

    
    \subsection{High Frequency Wafer Redesign}\label{sec:hf}

\begin{figure}[t]
        \centering
        \includegraphics[width=\linewidth]{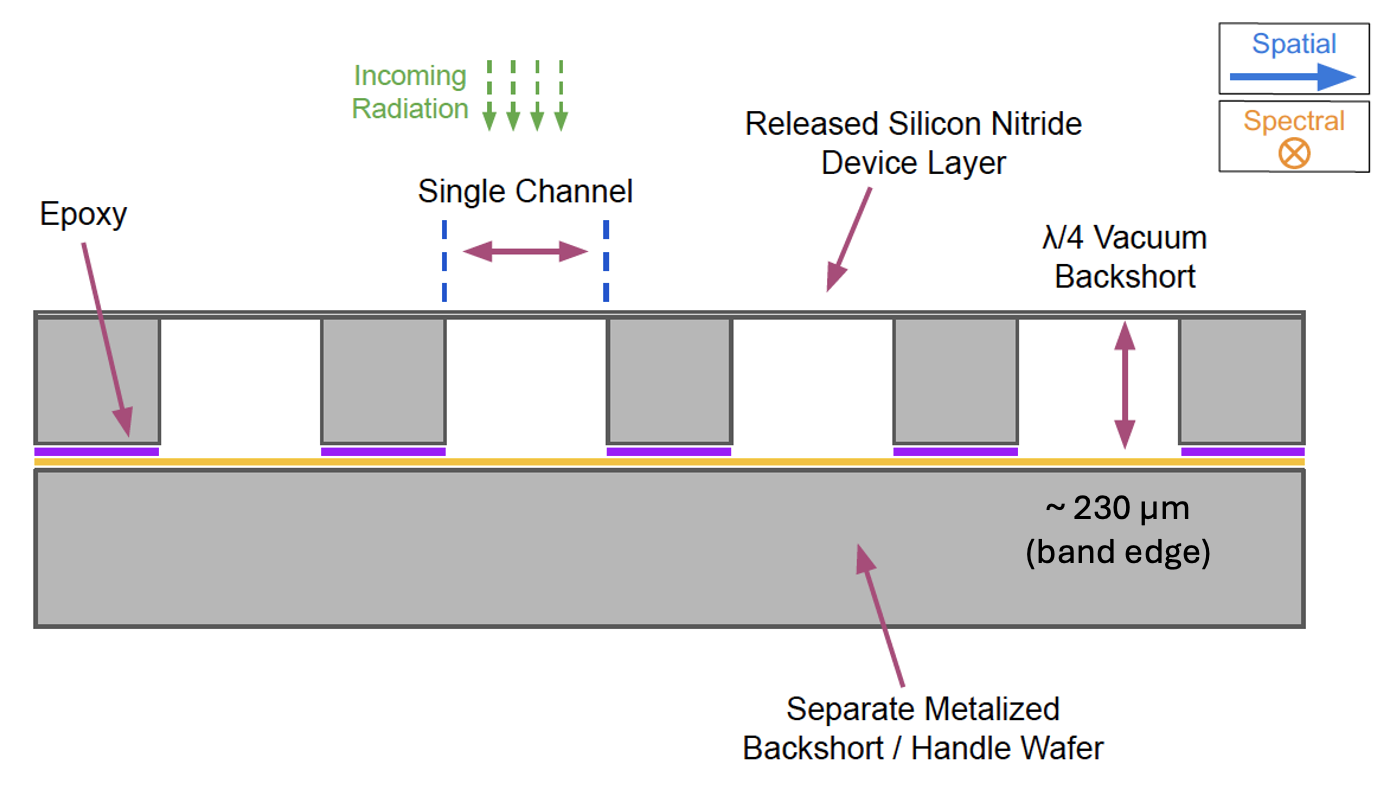}
        \caption{There is a wide spread in photon frequencies that each HF wafer is sensitive to. The new backshort design is optimized for the middle of the high frequency spectral bandpass of 230-325 GHz. Figure recreated with permission from the author \cite{Hunacek2020}.}
        \label{fig:backshort}
    \end{figure}

    Ideal detectors have a high fabrication yield, short time constants, and large optical efficiencies. These were some of the design parameters for the initial TIME detectors. One of the first designs had identical properties across LF and HF wafers, which increased the speed and yield of device fabrication. This first device was made with an etch-released silicon nitride device layer and a separate metalized backshort that were later epoxied together. Fig.~\ref{fig:backshort} shows an ideal TIME backshort, with lossless reflection of photons back into the absorber, with a distance defined by $(2n-1)\star \lambda/4$, for positive values of $n$. In this design, the backshort distance was set by the silicon device wafer thickness. The optimal $\lambda/4$ thickness for both the HF and LF wafers are slightly different, at 370$\mu$m and 420$\mu$m respectively. However, the extremely thin design of the HFs proved too fragile during the release process, resulting in a low yield. In order to improve yield and create interchangeable wafers between HF and LF modules, the backshort distance was fixed at 420$\mu$m.
    
    Laboratory testing of the thicker HF wafers showed long time constants and low optical efficiencies. This was partially expected from the LF-optimized backshort distance, however, the simulated impact to the optical efficiencies was only 10\%. The lab results put this closer to 20\%. In our updated design, the HF backshort distance was changed to its optimal value of 370$\mu$m which improved the optical efficiency. 
    
    Changes were also made to the etching process, since the long time constants were likely due to residual oxide left on the absorber. We changed to a wet release process using a buffered oxide etch and extended the etching period to 40 seconds. We are still waiting to test these newest HF wafers. 

\section{Future Observations}
    
    We plan to test the newly fabricated Kapton\texttrademark\ cables and HF detectors again at the ARO 12-meter Telescope on Kitt Peak, operated by the Arizona Radio Observatory (ARO). We anticipate significantly improved detector performance for the Winter 2025/26 season due to mechanical and thermal changes made to the cryostat.

    On the mechanical side, the spectrometer holder (shown in Figure~\ref{fig:spec}) was redesigned, and optical components for Spec B were added. We characterized changes to our optical path from these modifications through a mirror mapper designed by Mayer et al. (submitted to JATIS). This system samples the beam at multiple points in the optical chain, providing a detailed comparison to our optical models and helping to align the system more precisely.

    Additional internal improvements to the cryostat have reduced our cooldown time from a few weeks to a few days, allowing us to quickly characterize new detectors, as well as sustain a base temperature of 280 mK for up to 72 hours. This is due to replacing a mechanical heat switch between the 4K/50K and 1K stages with ultra-pure helium exchange gas, and improved $^{3}$He/$^{4}$He fridge cycling. Together, these changes enhance our ability to measure the instrument response to atmospheric and ground pickup and to increase the potential LIM field signal-to-noise.

    Previous TIME observations have characterized the spectro-imaging capabilities of the instrument through observations of local sources \cite{crites22}. At these redshifts, TIME is sensitive to a range of molecular emission lines, including $^{12}$CO(2--1), $^{13}$CO(2--1), H$^{12}$CN(3--2) and H$^{13}$CN(3--2). Early results from observations of the Central Molecular Zone (CMZ) of Sgr A, as well as other RF-bright sources like star forming region \texttt{G49.5} and carbon star \texttt{IRC +10 216}, were summarized in \cite{Butler2024}. This has motivated work to characterize and reduce both atmospheric noise through Principal Component Analysis, and inter-detector correlations through component masking. Yang et al. (in prep) will provide details of our end-to-end analysis pipeline on these sources. We anticipate this work will provide the framework needed to reduce our first CO/[CII] maps of the popular COSMOS field. 



\vfill

\section*{Acknowledgments}
The hardware in this work is supported by National Science Foundation award number 1910598 under the NSF-ATI program and the software and data analysis by 2308039 under the NSF-AAG program. Part of the research described in this proceeding was carried out at the Jet Propulsion Laboratory, California Institute of Technology, under a contract with the National Aeronautics and Space Administration. Author A. Crites also acknowledges support from NSF AST-1313319.
The TIME Collaboration also extends its gratitude towards the ARO 12m facility, operated by the University of Arizona and its employees, for supporting this project. We would also like to thank the Jet Propulsion Laboratory for allowing us access to its resources, and specifically the work of Anthony Turner.

\vfill

\end{document}